\documentclass{PoS}
\usepackage{amsmath}
\usepackage{amssymb}
\usepackage{graphicx}
\newcommand{\Phadr}{P_h}
\newcommand{\Pelec}{P_e}
\newcommand{\calF}{{\cal F}}

\definecolor{red}{rgb}{1,0,0}
\def\lesssim{\ \hbox{\raise 2pt \hbox{$<$} \kern -13pt
                     \lower 3pt \hbox{$\sim$}}\ }
\def\greatersim{\ \hbox{\raise 2pt \hbox{$>$} \kern -13pt
                     \lower 3pt \hbox{$\sim$}}\ }

\input epsf.tex
\def\desepsf(#1 width #2){\epsfxsize=#2 \epsfbox{#1}}

\def\katie{{\sc KaTie}}

\title{Small-$x$ Improved Transverse Momentum Dependent Factorization and some of its recent applications}

\ShortTitle{ITMD}

\author{\speaker{Krzysztof Kutak}\\
        Institute of Nuclear Physics Polish Academy of Sciences,\\
        Radzikowskiego 152, 31-342, Krakow\\
        E-mail: \email{krzysztof.kutak@ifj.edu.pl}}

\abstract{In this talk I am going to present some of recent phenomenological applications of the ITMD \cite{Kotko:2015ura,vanHameren:2016ftb} formalism extended to account for Sudakov form factor. In particular production of dijets in the future Electron Ion Collider is discussed. }

\FullConference{%
  *** The European Physical Society Conference on High Energy Physics (EPS-HEP2021), ***\\
  *** 26-30 July 2021 ***\\
  *** Online conference, jointly organized by Universität Hamburg and the research center DESY ***
}

\begin{document}
\section{Factorization for forward dijet production}
High energy collisions of protons and heavy nuclei at the Large Hadron Collider (LHC) provide a unique tool to probe
dense systems of quarks and gluons. Furthermore recently approved Electron Ion Collider \cite{AbdulKhalek:2021gbh} will allow for precision tomographic study of structure of 
hadrons ranging from proton to lead.
In particular interesting are processes where jets or particles are produced in the forward direction with respect to the probe which can be proton or electron. Kinematically, such final  states have large rapidities and therefore they trigger events in which the partons from the nucleus carry rather small longitudinal momentum fraction $x$.
This kinematic setup is perfectly suited to investigate the phenomenon of gluon saturation, which is expected to occur at some value of $x$ to prevent violation of the unitarity bound (for a review of this subject see Ref. \cite{Albacete:2014fwa}). The behavior of dense systems of partons when $x$ becomes small is predicted by  Quantum Chromodynamics (QCD) and leads to non-linear evolution equations known as B-JIMWLK \cite{Balitsky:1995ub,JalilianMarian:1997gr} equations (for review see \cite{Kovchegov:2012mbw,Gelis:2010nm}), which can be derived within the Color Glass Condensate (CGC) theory.
In CGC, the calculation of forward jet production 
relies on factorization, where  the dense target according to theoretical results is described with  nonlinear BK-JIMWLK equations.

The  description of multi-jet production is rather complicated~\cite{Marquet:2007vb}. A novel approach to such processes was initiated  in Ref.~\cite{Kotko:2015ura}. The framework is known as the small-$x$ Improved Transverse Momentum Dependent (ITMD) factorization. The ITMD formula accounts for: 
\begin{itemize}
\item complete kinematics of the scattering process with off-shell gluons, 
\item gauge invariant definitions of the TMD gluon densities, 
\item gauge invariant expressions for the off-shell hard matrix elements, 
\item it also recovers the high energy factorization (aka $k_T$-factorization)~\cite{Catani:1990eg,Deak:2009xt} in the limit of large off-shellness of the initial-state gluon from the nucleus and small-$x$ TMD factorization \cite{Dominguez:2011wm} in correlation limit i.e. when momenta of jets are much larger that momentum of off-shell gluon. 
\end{itemize}
\begin{figure}[t!]
  \begin{center}
    \includegraphics[width=0.99\textwidth]{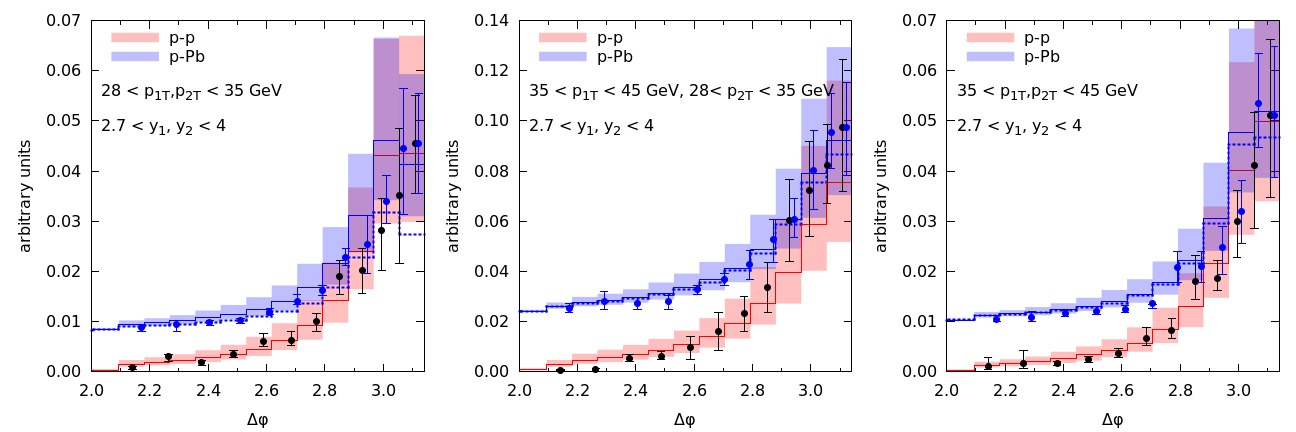}
  \end{center}
  \caption{
    Broadening of  azimuthal decorrelations in p-Pb collisions vs p-p collisions
    for different sets of cuts imposed on the jets' transverse momenta.
    The plots show  normalized cross sections as functions of the azimuthal distance between the two leading jets, $\Delta\phi$. 
    The points show the experimental data \cite{ATLAS:2019jgo} for p-p and p-Pb, where the p-Pb data were shifted by a pedestal, so that the values in the bin $\Delta\phi\sim \pi$ are the same.
    Theoretical calculations are represented by the histograms with uncertainty bands coming from varying the scale by factors 1/2 and 2.
  }
  \label{fig:broadening}
\end{figure}
Recently, the ITMD  factorization has been proved ~\cite{Altinoluk:2019fui}. Steps in further extension of the formalism to three and more jets were undertaken in Ref.~\cite{Bury:2018kvg,Bury:2020ndc}. While in the paper \cite{Altinoluk:2021ygv}
the ITMD was generalized to account for masses of produced hadrons as well as longitudinally polarized gluons (see also \cite{Boussarie:2020vzf} for discussion of the role of transverse gauge links).
Furthermore in the papers \cite{Fujii:2020bkl,Boussarie:2020pvm} the ITMD was shown to agree very well with full CGS results (accounting on top of kinematical twist for genuine twist effects) in the region dominated by hard jets i.e. $k_T,p_T> Q_s$
While the original ITMD formula, as well as the works studying the jet correlation limit within CGC, include gluon saturation effects, they do not account for all contributions proportional to logarithms of the hard scale set by the  large transverse momenta of jets -- the so-called Sudakov logarithms. It has been shown in Refs.~\cite{vanHameren:2014ala,vanHameren:2015uia} that inclusion of Sudakov logarithms is necessary in order to describe the LHC jet data at small $x$ but yet before the saturation regime. In the low $x$ domain, the resummation leading to the Sudakov logarithms has been developed in ~\cite{Mueller:2013wwa,Mueller:2012uf} see also \cite{Kutak:2014wga}. In the paper \cite{vanHameren:2019ysa}, it has been shown for the first time, that the interplay of saturation effects and the resummation of the Sudakov logarithms is essential to describe the small-$x$ forward-forward dijet data.
In this contribution we present two results that demonstrate relevance and importance of both effects i.e. nonlinearity accounting for saturation and Sudakov effects accounting for emission of soft gluons. 
\begin{itemize}
\item The first process under consideration is the inclusive dijet production
\begin{equation}
  \mathrm{p} \left(P_{\mathrm{p}}\right) + \mathrm{A} \left(P_{\mathrm{A}}\right) \to j_1 (p_1) + j_2 (p_2)+ X\ ,
\end{equation}
\item The second one is the dijet production in Deep Inelastic Scattering
\begin{equation}
  \mathrm{e} \left(P_{\mathrm{p}}\right) + \mathrm{A} \left(P_{\mathrm{A}}\right) \to j_1 (p_1) + j_2 (p_2)+ X\ ,
\end{equation}
\end{itemize}
where $A$ can be either the lead nucleus, or a proton. 
To describe the former of above processes, the hybrid approach has been used where one assumes that the proton $p$ is a dilute
projectile, whose partons are collinear to the beam and carry momenta $p=x_{\mathrm{p}} P_{\mathrm{p}}$.  
The hadron A is probed at a dense state. 
The jets $j_1$ and $j_2$ originate from hard partons produced in a collision of the probe $a$
with a gluon belonging to the dense system $A$. This gluon is off-shell, with momentum 
$k=x_{\mathrm{A}} P_{\mathrm{A}} + k_T$ and $k^2=-|\vec{k}_T|^2$.
The ITMD factorization formula (schematically) reads
\begin{equation}
d\sigma^{\mathrm{pA}\rightarrow j_1j_2+X}
=
\sum_{a,c,d} x_{\mathrm{p}} f_{a/\mathrm{p}}\left(x_{\mathrm{p}},\mu\right) \otimes
\sum_{i=1}^{2}\mathcal{K}_{ag^*\to cd}^{\left(i\right)}\left(q_T,k_T;\mu\right)
\otimes\Phi_{ag\rightarrow cd}^{\left(i\right)}\left(x_{\mathrm{A}},k_T,\mu\right)\,,
\label{eq:itmd}
\end{equation}   
The distributions $f_{{a/\mathrm{p}}}$ are the collinear PDFs corresponding to the large-$x$ gluons and quarks in the projectile. 
The functions $\mathcal{K}_{^{ag^*\to cd}}^{_{(i)}}$ are the hard matrix elements constructed from gauge-invariant off-shell amplitudes \cite{vanHameren:2012if}. 
The quantities $\Phi_{^{ag\rightarrow cd}}^{_{(i)}}$ are the TMD gluon distributions introduced in Ref.~\cite{Kotko:2015ura} and parametrise a dense state of the nucleus or the proton in terms of small-$x$ gluons.
Similarly the ITMD formula for e+A collision reads (once the contribution from longitudinally polarized gluons is neglected)
\begin{equation}
  d\sigma_{eh \to e' + 2j + X} =
  \frac{1}{4x\Pelec\!\cdot\!\Phadr}\,
\calF^{(3)}_{gg} (x, k_T, \mu)\,
\otimes
|\overline{M}_{eg^* \to e'+2j}|^2,
  \label{eq:ITMD}
\end{equation}
where $\calF^{(3)}_{gg} (x, k_T, \mu)$ is so called WW gluon density which has the interpretation of gluon number density.
To apply the ITMD formula one needs to construct the ITMD densities. One can use their operator definition and calculate their $x, k_T$ dependence using B-JIMWLK or using mean field approximation and express the ITMD distributions in terms of convolutions of dipole gluon density which is a solution of BK equation. We followed the second option and constructed the ITMD distributions from distributions given by the KS gluon density~\cite{Kutak:2012rf,vanHameren:2016ftb}.
\begin{figure}[t]
  \begin{center}
   \includegraphics[width=0.4\textwidth]{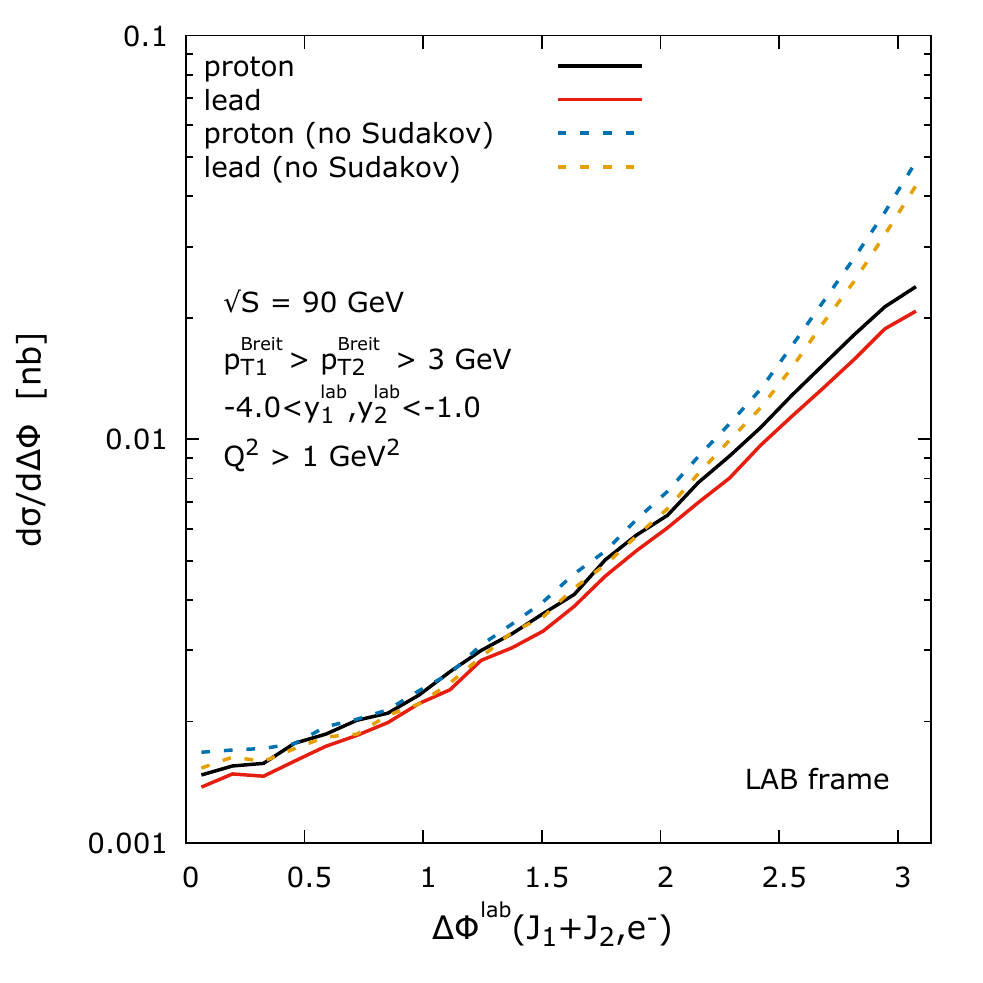}
   \includegraphics[width=0.4\textwidth]{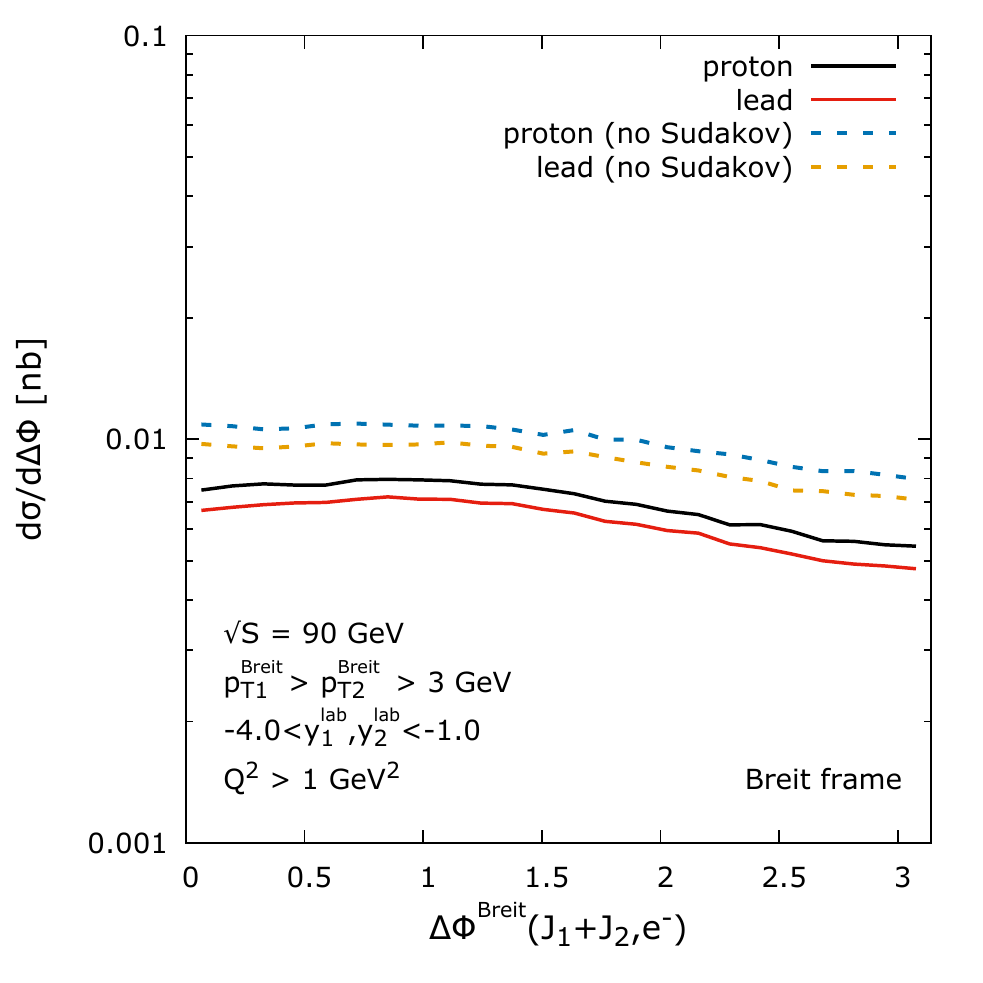}
  \end{center}
  \caption{Azimuthal correlations between the total transverse momentum of the
  dijet system and the transverse momentum of the scattered electron at EIC in
  two frames: the LAB frame (left), the Breit frame (right). The calculation has
  been done within the ITMD framework using {\katie} Monte Carlo \cite{vanHameren:2016kkz} with the Weizs\"acker-Williams gluon distribution obtained from the KS fit to HERA data.
  }
  \label{fig:angleplot}
\end{figure}
As it has been argued above the necessary additional element in order to provide realistic cross section is the Sudakov form factor. For the dijet production at the LHC we used DGLAP based Sudakov form factor \cite{vanHameren:2019ysa}. 
On fig.~\ref{fig:broadening}  we show normalized cross sections as functions of $\Delta\phi$ in p-p  and p-Pb collisions. 
The three panels correspond to three different cuts on the transverse momenta of the two leading jets.
The points with error bars represent experimental data from Ref.~\cite{ATLAS:2019jgo}. 
The main results for p-Pb collisions  are represented by blue solid lines in Fig.~\ref{fig:broadening}. 
The visible broadening comes from the interplay of the non-linear evolution of the initial state and the Sudakov resummation.
Similarly in the fig.~\ref{fig:angleplot} shows result for predictions for di-jet production at EIC obtained in \cite{vanHameren:2021sqc} for the cross section as a function of the angle between electron and dijet system. 
In this case we used Sudakov obtained in \cite{Mueller:2013wwa}.
A control result based on a calculation that neglects the Sudakov form factor is presented there as well.  The comparison of the two results clearly shows that while saturation effects are mild, the Sudakov effects are fairly large.  This feature is clearly visible in both the LAB and the Breit frame.  
\section{Conclusions}
In this contribution the ITMD factorization framework was presented. In particular I discussed its phenomenological applications to LHC and EIC physics. Furthermore it is clear that one needs to account for additional contributions coming from soft gluons and leading to Sudakov form factor. The combined contribution allows to describe shape of the LHC forward-forward dijet spectra. Using this knowledge we provide predictions for the EIC dijet crosssection. 
\section*{Acknowledgments}
The research is founded by Polish National Science Center grant no. DEC-2017/27/B/ST2/01985.

\end{document}